\documentclass[a4paper,10pt]{article}
\usepackage[utf8]{inputenc}
\usepackage[english]{babel}

\usepackage{amsmath}
\usepackage{amsthm}
\usepackage{amssymb}

\usepackage{authblk}
\usepackage{hyperref}
\usepackage{float}
\usepackage{graphicx}
\usepackage{listings}
\usepackage{etoolbox}
\usepackage{longtable}
\usepackage{multirow}
\usepackage[table,xcdraw]{xcolor}
\usepackage{ctable}
\usepackage{lscape}
\usepackage{enumitem}
\usepackage{wasysym}
\usepackage{placeins}
\usepackage{midpage}
\usepackage{diagbox}
\usepackage{caption}
\usepackage{subcaption}
\usepackage{algorithm}
\usepackage{algpseudocode}
\usepackage[bottom, hang]{footmisc}

\algnewcommand\algorithmicforeach{\textbf{for each}}
\algdef{S}[FOR]{ForEach}[1]{\algorithmicforeach\ #1\ \algorithmicdo}

\newtheorem{law}{Law}

\usepackage[backend=biber, sorting=nty]{biblatex}
\title{An Analysis of Quantum Annealing Algorithms for Solving the Maximum Clique Problem}
\author{Alessandro Gherardi}
\author{Alberto Leporati}
\affil{University of Milano-Bicocca \\
  Department of Informatics, Systems, and Communication\\
  Viale Sarca 336, Edificio U14, 20126 Milano, Italy \\
  \smallskip
  e-mail: a.gherardi9@campus.unimib.it \\
  \hspace{0.9cm}alberto.leporati@unimib.it
}
\date{}

\begin{document}
\maketitle

\begin{abstract}
Quantum annealers can be used to solve many (possibly NP-hard) combinatorial optimization problems, by formulating them as quadratic unconstrained binary optimization (QUBO) problems or, equivalently, using the Ising formulation.
In this paper we analyse the ability of quantum D-Wave annealers to find the maximum clique on a graph, expressed as a QUBO problem.
Due to the embedding limit of 164 nodes imposed by the anneler, we conducted a study on graph decomposition to enable instance embedding.
We thus propose a decomposition algorithm for the complementary maximum independent set problem, and a graph generation algorithm to control the number of nodes, the number of cliques, the density, the connectivity indices and the ratio of the solution size to the number of other nodes.
We then statistically analysed how these variables affect the quality of the solutions found by the quantum annealer.
The results of our investigation include recommendations on ratio and density limits not to be exceeded, as well as a series of precautions and a priori analyses to be carried out in order to maximise the probability of obtaining a solution close to the optimum.
\end{abstract}

\section{Introduction}

Since the appearance of the first quantum computers, the world of quantum computing has already achieved very good results in terms of performance in approaching certain NP-hard \cite{speedup1,speedup2} optimization problems.
A quantum CPU (QPU) that can be used to solve this kind of problems is the Advantage QPU~\cite{advantage} offered by D-Wave. This processor enables solutions to complex combinatorial problems to be computed using adiabatic quantum computation \cite{calc_quant_ad1, calc_quant_ad2} and quantum annealing, which searches for the lowest energy state of a target function by adiabatically evolving its states. The target function is called the Hamiltonian and is defined as \cite{dwave}:
\begin{equation}
    H_{ising}=-\frac{A(s)}{2} \left (\sum\limits_{i} \hat{\sigma}_{x}^{(i)}\right) + 
    \frac{B(s)}{2} \left (\sum\limits_{i} h_{i} \hat{\sigma}_{z}^{(i)} + \sum\limits_{i<j} J_{i,j} \hat{\sigma}_{z}^{(i)} \hat{\sigma}_{z}^{(j)} \right )
\end{equation}
where $\hat{\sigma}^{(i)}$ are the Pauli matrices operating on the qubits $q_{i}$, while $h_{i}$ and $J_{i,j}$ represent the qubit biases and coupling forces. The Hamiltonian of each problem can be embedded inside the QPU once the target problem has been reduced to either the Ising spin glass problem or the QUBO problem.

In this paper, we set out to analyse the solving capability of a D-Wave quantum computer on the NP-hard maximum clique problem (MC), which consists in finding a maximum subset of fully connected nodes in a graph.
The QUBO formulation of this problem, provided by \cite{QUBO}, can be written as:
\begin{equation}\label{eq:QUBO}
    H = -A\sum_{i=1}^Nx_i + B\sum_{(i,j)\in\Bar{E}}x_ix_j
\end{equation}
where the penalties $A$ and $B$ are set to 1 and 2 respectively, as explained in~\cite{ising}.
In particular, equation \eqref{eq:QUBO} refers to the QUBO function for the maximum independent set problem (IS), as it is simpler and more functional to implement.

Our aim is to understand which instances of the problem are harder to solve, what properties a suitable graph must have, and therefore what transformations and decompositions can be applied to obtain a suitable instance.
We then conducted a series of experiments to identify correlations between resolution quality and graph characteristics, with the goal of establishing a foundation for future research.
In what follows we will denote the undirected graphs as $G(V,E)$, where $V=\{1,\ldots,N\}$ is the set of $N$ vertices and $E$ is the set of edges.

The rest of the paper is organized as follows.
In Section \ref{sec:related-works} we present some related works taken from the literature.
In Section \ref{sec:methods-and-algorithms} we describe the mathematical methods and the software used in our investigation. In particular, we present our new graph decomposition and graph generation algorithms, and we describe the statistical tests we performed to analyse the solutions found by the quantum annealer on our benchmark graphs.
In Section \ref{sec:experimental-analysis} we describe the experiments we performed, and the corresponding results.
We also formulate two experimental laws, relating the clique size and the maximum independent set size with the quality of the solutions returned by the quantum annealers.
Finally, Section \ref{sec:conclusions} draws conclusions and outlines some directions for future work.

\section{Related Works\label{sec:related-works}}

Over time, various researches have been proposed to find the solution to the maximum clique problem, as well as various decomposition algorithms on graphs and methods to embed the problem inside a quantum computer. Most of the researches, like this paper, refer to the D-Wave \cite{dwave} hardware.

Quantum algorithms have already been benchmarked against classical heuristics, showing a significant improvement. In \cite{speedup1}, a speedup of about $10^8$ is observed in solving complex problems for instances optimised for D-Wave's Chimera architecture. \cite{speedup2} also uses D-wave hardware, but analyses quantum annealing models mathematically. It also demonstrates a speedup for several classes of problems by comparing a classical solver with a quantum solver and a simulated quantum solver. It is interesting to note that this speedup only applies to a few precise instances, suggesting that the problem may be due to a lack of the quantum hardware, which needs to be corrected and better calibrated. It should be noted that this paper dates back to 2014, and since then D-Wave has largely solved these problems.

In terms of problem formulation, the Ising format is provided by \cite{ising}, which defines several NP-hard problems in this format and is an essential starting point for embedding problems on quantum computers.
On the other hand, for the QUBO format there is no collection similar to \cite{ising}, so the definitions must be searched for in various articles, such as the one provided by \cite{QUBO} for the maximum clique problem.
After defining the problem, the researchers of this publication, analyzed the architecture of quantum computers D-Wave, understanding the need to decompose large problems.
They also analyzed various algorithms for this purpose, such as $k$-core, CH-partitioning, vertex splitting, and a combination of all three.
Finally, they study the differences between classical and quantum solvers, showing no significant differences between methods for small graphs, while for larger instances optimized for quantum computing the speedup compared to classical methods is important.

Regarding the specific problem of maximum clique, there is little strictly related work.
This is because the D-Wave technology is relatively new and constantly expanding, and the first computers allowed the inclusion of small problems (around 40-60 nodes) that could also be solved with classical algorithms.
A more modern example is presented in \cite{parallel}, which manages to process a graph with 120 nodes with acceptable results.
It does so by using the DBK decomposition algorithm and what researchers call parallel quantum computing. The latter method, as the name suggests, divides the problem into two subgraphs that are simultaneously embedded in different physical zones of the QPU (quantum CPU) and on different qubits, but solved independently. The advantage of this method lies in the possibility of avoiding the problem of the maximum size of the embedding.

Another way of looking at the clique problem is to reduce it to the maximum independent set problem. This problem is solved in \cite{Inde_set} using D-Wave technology. They also come to the same conclusions as above, showing an improvement in performance, but pointing out problems with the number of parameters required for optimisation, with the physical effects that limit the QPU, and the need to use optimised instances.

Given the problem of maximum embedding size, much research has been done on decomposing the graphs while keeping the maximum clique intact, in order to be able to include the problem instance in the D-Wave QPU. Various decompositions for many NP problems have already been proposed in \cite{dec_1985}. From there, many decomposition algorithms have been studied, leading to \cite{dec_new}, which specifically studies those mentioned by \cite{QUBO}. This research also introduces some very useful values used as upper and lower bounds to understand the size of a clique, and analyses how the choice of starting vertex can influence the decomposition algorithm.

However, much of this work does not consider the characteristics of a graph and how they may affect the solving process of the QPU, which we decided to test without relying on optimised instances or specific methods. In fact, many studies use ready-made instances for embedding in the Chimera or Pegasus topology of the QPU. We, on the other hand, want to understand in depth how the D-Wave solvers are affected by the structure of the base input without transformations, in order to add another level of detail or another point of view to the research already presented.

\section{Methods and Algorithms\label{sec:methods-and-algorithms}}

In this section we show how to embed instances of the maximum clique problem -- expressed in the QUBO formulation -- into the QPU and how to solve them using D-Wave solvers.
We also present a set of statistical methods for analysing the solutions, which are very useful for understanding how varying a feature of the input graph affects the quality of the results. Finally, we present two algorithms, one for decomposing a graph on the maximum independent set (IS) problem, and one for generating benchmark graphs. 

\subsection{D-wave Softwares}

D-Wave offers several tools for interfacing and programming on the QPU.
First and foremost is Leap\textsuperscript{\tiny{TM}} \cite{leap}, D-Wave's quantum cloud service, which allows users to interface with the QPU by sending problem instances to quantum computers, acting as a bridge between the IDE and the hardware.

Another service is the Solver API \cite{sapi}, called SAPI. This provides access to classical, quantum and hybrid solvers included in Leap\textsuperscript{\tiny{TM}} via access tokens.

D-Wave also offers a Python-based open source SDK named Ocean \cite{ocean}.
This includes everything needed for programming and a GitHub repository with various example user codes \cite{ocean_github}.
Among the libraries available are those dedicated to solvers, such as \textit{dimod} \cite{dimod}, which provides many functions dedicated to embedding problems.
Another very useful library in our case is \textit{dwave-networkx}~\cite{d_networkx}, an extension of the library \textit{networkx} \cite{networkx} which allows problems on graphs to be handled and solved using quantum methods.

Furthermore, D-Wave offers a variety of problem solving solutions. In addition to various classical solvers, hybrid and quantum solvers are also available. The latter differ in the type of problems they can accept: pure quantum solvers require problems in the form of Binary Quadratic Models (BQM) with binary variables and structured for the QPU, while hybrid solvers allow arbitrarily structured Quadratic Models (QM) with non-binary variables. Hybrid solvers use classical algorithms and hand off to quantum methods only those parts of the problems that can benefit most from them. Hybrid solvers include hybrid BQM solvers (accepting decision problems with a yes or no answer), hybrid CQM solvers (Constrained Quadratic Models, accepting constrained quadratic models with integer or binary variables, with possible constraints), and hybrid DQM solvers (Discrete Quadratic Models, accepting unconstrained and arbitrarily structured problems). These solvers are certainly useful and very powerful, but in our case we decided to focus only on quantum solvers in order to better understand their structure and solving capabilities. 

Speaking of quantum solvers, we tested two for our work: \textit{DWaveSampler} \cite{sampler} and \textit{DWaveCliqueSampler} \cite{clique_sampler}. The latter is a specific solver for the maximum clique problem that needs nothing else to work, while \textit{DWaveSampler} is the basic sampler that needs to be integrated into the QPU before returning the desired solution. This step is a bit more complex, but allows for different ways of embedding the problem instance into the QPU, giving more control. In our case, we tested several embedding functions provided by D-Wave. The best one was \textit{AutoEmbeddingComposite}, which sends the problem in its native form, i.e. as a binary quadratic model, and only applies embedding if necessary. This embedding involves finding the best qubit configuration on the QPU and the most efficient link chain to process the instance as efficiently as possible for the processor.

Finally, an important factor to consider is the fact that solvers are currently able to incorporate graphs of up to 164 nodes into the QPU, which are larger than those previously tested. Although a graph of this size can still be processed by a classical algorithm in a reasonable time, it allows us to analyse the potential of quantum methods better than in the past, and to estimate how the solving power might evolve on graphs of larger size.

\subsection{Statistical Methods}

In order to analyse the results obtained from the experiments, we decided to use some statistical tests to understand whether by changing a parameter the groups of results are significantly different. These are introduced and used to analyse the relationship between the extracted clique size and the original clique size when varying a chosen parameter. All proposed tests assume that there are $k$ groups of values, divided into $n$ samples per group.

\textbf{Cochran's test:} this statistical test, proposed in 1967 in \cite{cochran}, measures whether the variances of groups are homogeneous.

\textbf{Shapiro-Wilk's test:} this test, explained in \cite{shapiro}, is one of the most useful for measuring the normality of a data set. It is tested by comparing two alternative variance estimators, one non-parametric and one parametric.

\textbf{Test ANOVA:} Analysis of Variance, also known as ANOVA, is a statistical technique for analysing multiple data sets by comparing within-group variability (MQE) with between-group variability (MQF). It should be noted that ANOVA requires homoscedasticity and a normal trend for each data set. This last point will be important in our case, as the data sets under consideration will not all have normal trends. However, it should be noted that in \cite{AnovaNonNorm} it is shown that ANOVA analysis is not overly affected by the lack of normality of the data, in fact it remains a valid tool in most cases. The value $F=MQF/MQE$ is calculated and compared with a tabulated value; if $F$ is lower, it can be concluded that there are no significant differences between the groups studied. Otherwise, the LDS (least significant difference) test should be performed.

\textbf{Test LDS}: The LDS value is calculated and for each pair of groups the differences between the means in absolute value are calculated. A difference greater than the LDS value indicates significant differences in the pair examined, allowing conclusions to be drawn based on the context of the study.

\textbf{Kruskal-Wallis's test}: This test, explained in \cite{kruskal-wallis}, is the non-parametric equivalent of ANOVA. It analyses more than two groups and does not require normality of the data, only that all sets have a similar distribution, such as the same asymmetry to the right or left. The sum of the ranks $R_i$ for each group is calculated; if the initial hypothesis is true, the $R_i$ should have a similar value. From here, the $H$ statistic is calculated and compared to a tabulated value, and if $H$ is smaller, we can conclude that the initial hypothesis is accepted. Otherwise, as with ANOVA, it is necessary to perform further tests to find out which groups of data do not belong to the population being studied, and thus which groups are significantly different from the others.

\textbf{Mann-Whitney test} This method, explained in \cite{mann-whitney}, is another non-parametric test to analyse whether or not certain data belong to the same population and is carried out after rejecting the initial Kruskal-Wallis hypothesis. Groups are paired, and rank sums and then the U statistic are calculated for each. The value $U$ is compared to a tabulated value and if it is greater, it can be concluded that the two groups belong to the same population and therefore there are no significant differences.
\subsection{IS-decomposition}
In order to address the issue of the maximum embedding size, we have developed a novel decomposition algorithm that is specifically designed to tackle the maximum independent set problem. This is because the quantum algorithms that we have employed are capable of solving this problem.

A very important preliminary factor is the definition of an upper bound on the size of the clique to be sought. We have chosen to use the \textit{chromatic number}, as mentioned in \cite{dec_new}. The difficulty lies in the fact that finding the chromatic number is an NP-hard problem. However, there are greedy colouring algorithms that allow us to obtain a good approximation of this value.

For the \textit{IS-decomposition} algorithm, we started with the idea that if we choose three vertices, and two of them are connected to the third but not to each other, then removing the central vertex would result in two unconnected vertices that could be part of the maximum independent set. Another consideration is that vertices with a higher degree have a lower probability of being part of the independent set, so the search for vertices to remove starts from the vertices with the highest degree.

The algorithm takes as input the complementary graph to the one containing the maximum clique and removes nodes until it reaches the size specified by the user, or until the removal does not reduce the value of the colour number on the original graph below a specified threshold. It should be noted that choosing too low a value can have destructive effects on the maximum independent set.

Algorithm \ref{alg:is-dec} shows the pseudocode of the method.  It is subject to several control conditions: the {\bf while} loop is executed at most $2n$ times, and the function $get\_max\_degree\_vertex\_to\_remove$ recursively searches for a triple of vertices satisfying the initial condition at most 5 times.
The $get\_random\_vertex$ function also searches for a removable vertex, but with randomly chosen triples, and stops after a given number of iterations without a result. Furthermore, the vertices connected to the removed vertex are placed in a list of non-removable vertices in order to retain the possible candidates for the independent set.

It is very difficult to formally prove that the algorithm works in all cases. However, our empirical findings indicate that in the majority of cases, the maximum independent set (or one of the possible ones if there are more) is preserved.
Additionally, the time complexity for the worst case scenario, where a fully connected graph is provided as input, is $O(n^5)$. However, it is very unlikely that the worst case will occur by chance, as the user would not be expected to attempt to decompose a ready-made clique; hence the algorithm should run in a significantly shorter time than the worst case.

\begin{figure}[t]
    \centering
    \includegraphics[width=0.8\textwidth]{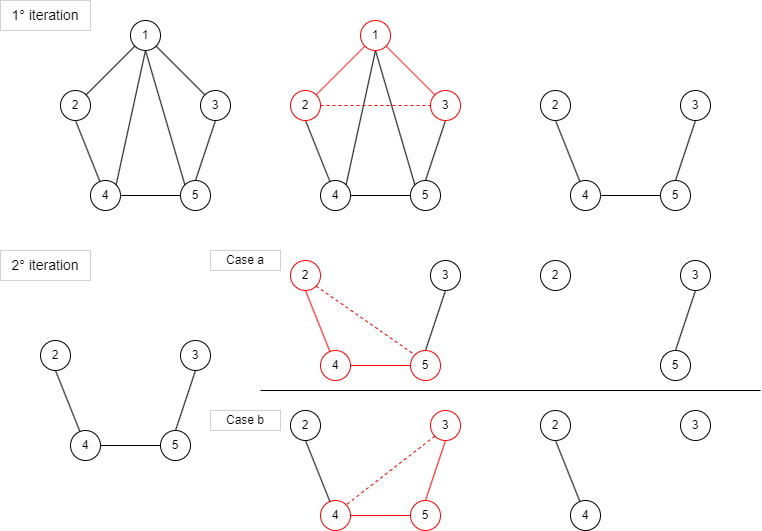}
    \caption{Working example of the IS-decomposition algorithm}
    \label{fig:is-dec}
\end{figure}

An illustrative example of the algorithm's operation is presented in Figure~\ref{fig:is-dec}. It can be observed that at the initial iteration, the triple [1,2,3] is selected, resulting in the elimination of vertex 1 (having the highest degree). A comparable outcome is observed for the valid triples [1,2,5] and [1,3,4]. In the event that elimination was not possible, a random vertex between 4 and 5 would have been selected, as it is second in degree. At the second iteration, the algorithm makes a random decision to eliminate vertex 4 or 5, which are equal in degree and both have neighbours that satisfy the initial condition. In both cases, one of the possible independent sets is retained. Upon completion of the second iteration, no vertex satisfies the specified restrictions for removal, and the algorithm terminates. In fact, no independent set is identified. However, the size of the initial graph has been reduced in order to facilitate subsequent computations of the solution.

\begin{algorithm}[t]
    \caption{IS-decomposition}\label{alg:is-dec}
    \begin{algorithmic}
        \Function{decompose\_IS}{$G$, final\_dim, min\_cn}
            \While {dim$(G)$ $>$ dim\_original}
                \State $v,v',v''\leftarrow$ get\_max\_degree\_vertex\_to\_remove($G$, min\_cn)
                \If {$v$ not found}
                    \State $v,v',v'' \leftarrow$ get\_random\_vertex($G$, min\_cn) \Comment{If $v$ satisfies \\ \hfill requirements}
                \EndIf
                \If {$v$ found} \Comment{If $v$ found in either case}
                \If {check\_if\_$v$\_can\_be\_removed($G$, $v$, min\_cn)}
                    \State $G$.remove($v$)
                    \State Add\_vertices\_to\_non\_removing\_list($L, v',v''$)
                \EndIf 
                \Else
                \State Stop \Comment{If $v$ not found, the function stops}
                \EndIf
                \EndWhile
        \EndFunction
    \end{algorithmic}
\end{algorithm}

\subsection{Algorithm for Creating Benchmark Graphs}

Another challenge we encountered was the difficulty of locating standard graphs, or at the very least, those whose maximum clique nodes were known.
To address this issue, we developed a highly controllable graph creation algorithm with the objective of obtaining all the necessary information for subsequent statistical investigations.
The algorithm permits the user to select the number of clique nodes (n\_node), the number of cliques (n\_cli) and the number of nodes not belonging to cliques (ex\_node).
Furthermore, two Boolean variables may be set to indicate whether arcs should be added between the cliques (add\_edges) and whether the size of the cliques should be randomised (rand\_cli), keeping the variable n\_node as the maximum value.
The method returns the created graph, along with a list of nodes belonging to the created cliques.
Algorithm~\ref{alg:graphCreation} illustrates the key steps of the method.
\begin{algorithm}[H]
    \caption{Graph-creation}\label{alg:graphCreation}
    \begin{algorithmic}
        \Function{graph\_creation}{n\_node, ex\_node, n\_cli, add\_edges, rand\_cli}
        \State final\_dim $\leftarrow$ n\_node$\cdot$n\_cli+ex\_node
        \State $g$, cli\_list, ex\_node\_list $\leftarrow$ graph\_with\_cliques(final\_dim, n\_node, n\_cli, \\
        \hspace{7.45cm}rand\_cli)
        \ForEach {$v$ $\in$ ex\_nodes\_list}
            \State $g$ $\leftarrow$ add\_edges\_from\_node($v$, num\_of\_random\_links)
        \EndFor
        \If{add\_edges}
            \State $g$ $\leftarrow$ add\_edges\_between\_cliques($g$, num\_of\_random\_links)
        \EndIf
        \State \Return $g$, cli\_list
        \EndFunction        
    \end{algorithmic}
\end{algorithm}
The algorithm exhibits a high degree of randomness. The nodes of the cliques and the number of arcs created are selected at random. In particular, for the arcs, each addition is applied only if it does not increase the value of the chromatic number calculated on the initial graph with only the cliques. This ensures that the original size of the maximum clique inserted remains intact. Finally, the percentage determining the number of arcs in the two functions can be set, thus enabling the creation of graphs with varying densities, which can then be analysed in order to ascertain the impact of this factor.

The algorithm was thus created with the objective of achieving a high degree of variability between the graphs generated, thereby enabling the creation of instances that accurately reflect the search space and from which to extract results that are as descriptive as possible of the resolving capacity of the QPU. In addition, the aim was to minimise the possibility of having chosen graphs that could have introduced a bias into the analysis of the results.

\section{Experimental Analysis\label{sec:experimental-analysis}}

This section presents the results of the various tests conducted. The objective is to idenitfy relationships between the structure of a graph and the ability of quantum methods to find the max clique. All experiments are structured in four steps: creation of $n$ graphs with different characteristics, calculation of the maximum clique, elimination of null results with the addition of new instances to compensate (if necessary) and, finally, statistical investigation of the results.

Once the four steps have been completed, if the statistical information is not deemed relevant, new instances are added to those already generated, and this process is repeated until a statistical trend emerges. Furthermore, this step elucidates the rationale behind the differing number of graphs analysed for each experiment.

Furthermore, it should be noted that all the tests conducted were independent of one another. This was achieved by creating new instances for each experiment, thereby allowing for a comprehensive analysis of the problem.

\subsection{Initial Exploratory Analysis}

A preliminary analysis was conducted to identify the most pertinent variables for achieving optimal results with quantum solvers and, thus, understand how to structure subsequent experiments.

Approximately 100 graphs were created, divided into five groups of varying sizes. The graphs were not generated simultaneously; instead, the results of each group influenced the creation of subsequent ones.

The final size of the graphs exhibited considerable variability, with a majority falling within the range of 90 to 100 nodes. This observation was made due to the tendency of the solvers to consistently report optimal solutions for smaller values, and conversely, suboptimal solutions for larger values. The identified range, however, encompasses heterogeneous results that are more useful for analysis. With regard to the size of the cliques, the values observed in the range 5-80 were found to be quite disparate. This is because the various results reported heterogeneity in the size returned. Furthermore, the number of cliques does not appear to be a determining factor in influencing the resolving quality of the method.

These initial experiments demonstrated the necessity of generating graphs of a total size close to 100 nodes in order to understand the behaviour of the quantum solver in a situation appropriate to the current capacity of the method. Furthermore, we observed a correlation between clique size and the quality of the results, necessitating further investigation. Based on these two conclusions, we structured the subsequent tests.

\subsection{Ratio Between Clique Size and External Nodes}

We begin our analysis with this experiment, as it yields the most relevant outcome for our research and for the other experiments. Our objective is to ascertain the degree to which the size of the clique is a significant factor for the quantum method.

A total of 42 graphs have been generated, with a final size of 100 nodes and the number of cliques at 1. These have been divided into seven different groups by the ratio of the number of clique nodes to the number of external nodes. The maximum cliques have been calculated by the D-Wave solver, with instances that produce a null output being replaced in order to avoid the introduction of staggered data.

Firstly, we calculated the ratio between the clique size returned by the D-Wave solver and the original clique size for each graph. On this data, coupled with the clique/outer node ratio, we performed some statistical tests. Initially, we performed the Cochran test \cite{cochran} with a positive result, and then proceeded to perform the Shapiro-Wilk test \cite{shapiro}. Normality is only followed for groups with lower ratios, but thanks to \cite{AnovaNonNorm} we nevertheless felt entitled to perform an ANOVA test, in conjunction with the Kruskal-Wallis test \cite{kruskal-wallis} for additional security (all data have left skewness). Both tests rejected the initial hypothesis that there were no differences between the groups. Further tests were required to ascertain which groups differed from the others. These included the LDS post-ANOVA and the Mann-Whitney test post-Kruskal-Wallis. In particular, various levels of significance $\alpha$ were used for the latter test to identify which groups were truly different from the others. The results are presented in  Table \ref{tab:mds_mann}. It can be observed that the comparisons between the pairs of groups demonstrate how group 7 (ratio = 0.42) differs from the majority of the others, in particular from group 1, which could be considered as a reference.

This result represents the most significant outcome of our investigation. The limiting ratio of 0.42, or equivalently a clique of 30 nodes on a graph of 100 total, marks the point at which the quantum solver begins to exhibit a decline in resolution quality.

From this result, we formulated the following two \emph{laws}, that is, mathematical relations extrapolated from empirical data and capable of explaining an experimental observation with a sufficient degree of precision. It should be noted that these laws are by no way formally proven theorems.
\begin{law}
    For the maximum clique problem, for the same final dimension of the graph, as the clique size increases, and thus the size of the maximum independent set on the complementary graph, the solution returned by the quantum solver approaches the optimum.
\end{law}
The same concept applies to the minimum-coverage problem, which is complementary to IS on the same graph:
\begin{law}
    For the minimum vertex cover problem, for the same final graph size, as the cover size increases, and thus as the size of the maximum independent set on the same graph decreases, the solution returned by the quantum solver deviates from the optimum solution.
\end{law}

\begin{table}[t]
\centering
\resizebox{0.9\columnwidth}{!}{%
\begin{tabular}{ccccccccccccccc}
\cline{1-7} \cline{9-15}
\multicolumn{1}{|c|}{MDS}           & \multicolumn{1}{c|}{G2}                                  & \multicolumn{1}{c|}{G3}                                  & \multicolumn{1}{c|}{G4}                                  & \multicolumn{1}{c|}{G5}                                  & \multicolumn{1}{c|}{G6}                                  & \multicolumn{1}{c|}{G7}                                  & \multicolumn{1}{c|}{} & \multicolumn{1}{c|}{MW $\alpha$=0,05} & \multicolumn{1}{c|}{G2}                                  & \multicolumn{1}{c|}{G3}                                  & \multicolumn{1}{c|}{G4}                                  & \multicolumn{1}{c|}{G5}                                  & \multicolumn{1}{c|}{G6}                                  & \multicolumn{1}{c|}{G7}                                  \\ \cline{1-7} \cline{9-15} 
\multicolumn{1}{|c|}{G1}              & \multicolumn{1}{l|}{{\color[HTML]{009901} $\checkmark$}} & \multicolumn{1}{c|}{{\color[HTML]{009901} $\checkmark$}} & \multicolumn{1}{c|}{{\color[HTML]{009901} $\checkmark$}} & \multicolumn{1}{c|}{{\color[HTML]{009901} $\checkmark$}} & \multicolumn{1}{c|}{{\color[HTML]{FF0000} X}}            & \multicolumn{1}{c|}{{\color[HTML]{FF0000} X}}            & \multicolumn{1}{c|}{} & \multicolumn{1}{c|}{G1}               & \multicolumn{1}{l|}{{\color[HTML]{009901} $\checkmark$}} & \multicolumn{1}{l|}{{\color[HTML]{009901} $\checkmark$}} & \multicolumn{1}{c|}{{\color[HTML]{FF0000} X}}            & \multicolumn{1}{c|}{{\color[HTML]{FF0000} X}}            & \multicolumn{1}{c|}{{\color[HTML]{FF0000} X}}            & \multicolumn{1}{c|}{{\color[HTML]{FF0000} X}}            \\ \cline{1-7} \cline{9-15} 
\multicolumn{1}{|c|}{G2}              & \multicolumn{1}{c|}{}                                    & \multicolumn{1}{c|}{{\color[HTML]{009901} $\checkmark$}} & \multicolumn{1}{c|}{{\color[HTML]{009901} $\checkmark$}} & \multicolumn{1}{c|}{{\color[HTML]{009901} $\checkmark$}} & \multicolumn{1}{c|}{{\color[HTML]{009901} $\checkmark$}} & \multicolumn{1}{c|}{{\color[HTML]{FF0000} X}}            & \multicolumn{1}{c|}{} & \multicolumn{1}{c|}{G2}               & \multicolumn{1}{c|}{}                                    & \multicolumn{1}{l|}{{\color[HTML]{009901} $\checkmark$}} & \multicolumn{1}{l|}{{\color[HTML]{009901} $\checkmark$}} & \multicolumn{1}{l|}{{\color[HTML]{009901} $\checkmark$}} & \multicolumn{1}{l|}{{\color[HTML]{009901} $\checkmark$}} & \multicolumn{1}{l|}{{\color[HTML]{009901} $\checkmark$}} \\ \cline{1-7} \cline{9-15} 
\multicolumn{1}{|c|}{G3}              & \multicolumn{1}{c|}{}                                    & \multicolumn{1}{c|}{}                                    & \multicolumn{1}{l|}{{\color[HTML]{009901} $\checkmark$}} & \multicolumn{1}{l|}{{\color[HTML]{009901} $\checkmark$}} & \multicolumn{1}{l|}{{\color[HTML]{009901} $\checkmark$}} & \multicolumn{1}{c|}{{\color[HTML]{FF0000} X}}            & \multicolumn{1}{c|}{} & \multicolumn{1}{c|}{G3}               & \multicolumn{1}{c|}{}                                    & \multicolumn{1}{c|}{}                                    & \multicolumn{1}{l|}{{\color[HTML]{009901} $\checkmark$}} & \multicolumn{1}{l|}{{\color[HTML]{009901} $\checkmark$}} & \multicolumn{1}{l|}{{\color[HTML]{009901} $\checkmark$}} & \multicolumn{1}{c|}{{\color[HTML]{FF0000} X}}            \\ \cline{1-7} \cline{9-15} 
\multicolumn{1}{|c|}{G4}              & \multicolumn{1}{c|}{}                                    & \multicolumn{1}{c|}{}                                    & \multicolumn{1}{c|}{}                                    & \multicolumn{1}{l|}{{\color[HTML]{009901} $\checkmark$}} & \multicolumn{1}{l|}{{\color[HTML]{009901} $\checkmark$}} & \multicolumn{1}{c|}{{\color[HTML]{FF0000} X}}            & \multicolumn{1}{c|}{} & \multicolumn{1}{c|}{G4}               & \multicolumn{1}{c|}{}                                    & \multicolumn{1}{c|}{}                                    & \multicolumn{1}{c|}{}                                    & \multicolumn{1}{l|}{{\color[HTML]{009901} $\checkmark$}} & \multicolumn{1}{l|}{{\color[HTML]{009901} $\checkmark$}} & \multicolumn{1}{c|}{{\color[HTML]{FF0000} X}}            \\ \cline{1-7} \cline{9-15} 
\multicolumn{1}{|c|}{G5}              & \multicolumn{1}{c|}{}                                    & \multicolumn{1}{c|}{}                                    & \multicolumn{1}{c|}{{\color[HTML]{009901} }}             & \multicolumn{1}{c|}{}                                    & \multicolumn{1}{l|}{{\color[HTML]{009901} $\checkmark$}} & \multicolumn{1}{l|}{{\color[HTML]{009901} $\checkmark$}} & \multicolumn{1}{c|}{} & \multicolumn{1}{c|}{G5}               & \multicolumn{1}{c|}{}                                    & \multicolumn{1}{c|}{}                                    & \multicolumn{1}{c|}{}                                    & \multicolumn{1}{c|}{}                                    & \multicolumn{1}{l|}{{\color[HTML]{009901} $\checkmark$}} & \multicolumn{1}{l|}{{\color[HTML]{009901} $\checkmark$}} \\ \cline{1-7} \cline{9-15} 
\multicolumn{1}{|c|}{G6}              & \multicolumn{1}{c|}{}                                    & \multicolumn{1}{c|}{}                                    & \multicolumn{1}{c|}{}                                    & \multicolumn{1}{c|}{}                                    & \multicolumn{1}{c|}{}                                    & \multicolumn{1}{l|}{{\color[HTML]{009901} $\checkmark$}} & \multicolumn{1}{c|}{} & \multicolumn{1}{c|}{G6}               & \multicolumn{1}{c|}{}                                    & \multicolumn{1}{c|}{}                                    & \multicolumn{1}{c|}{}                                    & \multicolumn{1}{c|}{}                                    & \multicolumn{1}{c|}{}                                    & \multicolumn{1}{l|}{{\color[HTML]{009901} $\checkmark$}} \\ \cline{1-7} \cline{9-15} 
                                      &                                                          &                                                          &                                                          &                                                          &                                                          &                                                          &                       &                                       &                                                          &                                                          &                                                          &                                                          &                                                          &                                                          \\ \cline{1-7} \cline{9-15} 
\multicolumn{1}{|c|}{MW $\alpha$=0,1} & \multicolumn{1}{c|}{G2}                                  & \multicolumn{1}{c|}{G3}                                  & \multicolumn{1}{c|}{G4}                                  & \multicolumn{1}{c|}{G5}                                  & \multicolumn{1}{c|}{G6}                                  & \multicolumn{1}{c|}{G7}                                  & \multicolumn{1}{c|}{} & \multicolumn{1}{c|}{MW $\alpha$=0,2}  & \multicolumn{1}{c|}{G2}                                  & \multicolumn{1}{c|}{G3}                                  & \multicolumn{1}{c|}{G4}                                  & \multicolumn{1}{c|}{G5}                                  & \multicolumn{1}{c|}{G6}                                  & \multicolumn{1}{c|}{G7}                                  \\ \cline{1-7} \cline{9-15} 
\multicolumn{1}{|c|}{G1}              & \multicolumn{1}{l|}{{\color[HTML]{009901} $\checkmark$}} & \multicolumn{1}{c|}{{\color[HTML]{009901} $\checkmark$}} & \multicolumn{1}{c|}{{\color[HTML]{FF0000} X}}            & \multicolumn{1}{c|}{{\color[HTML]{FF0000} X}}            & \multicolumn{1}{c|}{{\color[HTML]{FF0000} X}}            & \multicolumn{1}{c|}{{\color[HTML]{FF0000} X}}            & \multicolumn{1}{c|}{} & \multicolumn{1}{c|}{G1}               & \multicolumn{1}{c|}{{\color[HTML]{009901} $\checkmark$}} & \multicolumn{1}{c|}{{\color[HTML]{009901} $\checkmark$}} & \multicolumn{1}{c|}{{\color[HTML]{FF0000} X}}            & \multicolumn{1}{c|}{{\color[HTML]{FF0000} X}}            & \multicolumn{1}{c|}{{\color[HTML]{FF0000} X}}            & \multicolumn{1}{c|}{{\color[HTML]{FF0000} X}}            \\ \cline{1-7} \cline{9-15} 
\multicolumn{1}{|c|}{G2}              & \multicolumn{1}{c|}{}                                    & \multicolumn{1}{c|}{{\color[HTML]{009901} $\checkmark$}} & \multicolumn{1}{c|}{{\color[HTML]{009901} $\checkmark$}} & \multicolumn{1}{c|}{{\color[HTML]{009901} $\checkmark$}} & \multicolumn{1}{c|}{{\color[HTML]{009901} $\checkmark$}} & \multicolumn{1}{c|}{{\color[HTML]{FF0000} X}}            & \multicolumn{1}{c|}{} & \multicolumn{1}{c|}{G2}               & \multicolumn{1}{c|}{}                                    & \multicolumn{1}{c|}{{\color[HTML]{009901} $\checkmark$}} & \multicolumn{1}{c|}{{\color[HTML]{009901} $\checkmark$}} & \multicolumn{1}{c|}{{\color[HTML]{009901} $\checkmark$}} & \multicolumn{1}{c|}{{\color[HTML]{009901} $\checkmark$}} & \multicolumn{1}{c|}{{\color[HTML]{FF0000} X}}            \\ \cline{1-7} \cline{9-15} 
\multicolumn{1}{|c|}{G3}              & \multicolumn{1}{c|}{}                                    & \multicolumn{1}{c|}{}                                    & \multicolumn{1}{c|}{{\color[HTML]{009901} $\checkmark$}} & \multicolumn{1}{c|}{{\color[HTML]{009901} $\checkmark$}} & \multicolumn{1}{c|}{{\color[HTML]{009901} $\checkmark$}} & \multicolumn{1}{c|}{{\color[HTML]{FF0000} X}}            & \multicolumn{1}{c|}{} & \multicolumn{1}{c|}{G3}               & \multicolumn{1}{c|}{}                                    & \multicolumn{1}{c|}{}                                    & \multicolumn{1}{c|}{{\color[HTML]{009901} $\checkmark$}} & \multicolumn{1}{c|}{{\color[HTML]{009901} $\checkmark$}} & \multicolumn{1}{c|}{{\color[HTML]{009901} $\checkmark$}} & \multicolumn{1}{c|}{{\color[HTML]{FF0000} X}}            \\ \cline{1-7} \cline{9-15} 
\multicolumn{1}{|c|}{G4}              & \multicolumn{1}{c|}{}                                    & \multicolumn{1}{c|}{}                                    & \multicolumn{1}{c|}{}                                    & \multicolumn{1}{c|}{{\color[HTML]{009901} $\checkmark$}} & \multicolumn{1}{c|}{{\color[HTML]{009901} $\checkmark$}} & \multicolumn{1}{c|}{{\color[HTML]{FF0000} X}}            & \multicolumn{1}{c|}{} & \multicolumn{1}{c|}{G4}               & \multicolumn{1}{c|}{}                                    & \multicolumn{1}{c|}{}                                    & \multicolumn{1}{c|}{}                                    & \multicolumn{1}{c|}{{\color[HTML]{009901} $\checkmark$}} & \multicolumn{1}{c|}{{\color[HTML]{009901} $\checkmark$}} & \multicolumn{1}{c|}{{\color[HTML]{FF0000} X}}            \\ \cline{1-7} \cline{9-15} 
\multicolumn{1}{|c|}{G5}              & \multicolumn{1}{c|}{}                                    & \multicolumn{1}{c|}{}                                    & \multicolumn{1}{c|}{}                                    & \multicolumn{1}{c|}{}                                    & \multicolumn{1}{c|}{{\color[HTML]{009901} $\checkmark$}} & \multicolumn{1}{c|}{{\color[HTML]{FF0000} X}}            & \multicolumn{1}{c|}{} & \multicolumn{1}{c|}{G5}               & \multicolumn{1}{c|}{}                                    & \multicolumn{1}{c|}{}                                    & \multicolumn{1}{c|}{}                                    & \multicolumn{1}{c|}{}                                    & \multicolumn{1}{c|}{{\color[HTML]{009901} $\checkmark$}} & \multicolumn{1}{c|}{{\color[HTML]{FF0000} X}}            \\ \cline{1-7} \cline{9-15} 
\multicolumn{1}{|c|}{G6}              & \multicolumn{1}{c|}{}                                    & \multicolumn{1}{c|}{}                                    & \multicolumn{1}{c|}{}                                    & \multicolumn{1}{c|}{}                                    & \multicolumn{1}{c|}{}                                    & \multicolumn{1}{c|}{{\color[HTML]{009901} $\checkmark$}} & \multicolumn{1}{c|}{} & \multicolumn{1}{c|}{G6}               & \multicolumn{1}{c|}{}                                    & \multicolumn{1}{c|}{}                                    & \multicolumn{1}{c|}{}                                    & \multicolumn{1}{c|}{}                                    & \multicolumn{1}{c|}{}                                    & \multicolumn{1}{c|}{{\color[HTML]{009901} $\checkmark$}} \\ \cline{1-7} \cline{9-15} 
\end{tabular}%
}
\captionsetup{justification=centering}
\caption{Groups divided by the different tests, $\checkmark$ if the two groups show no significant differences, X otherwise}
\label{tab:mds_mann}
\end{table}

The reason for this phenomenon can be attributed to the number of nodes that are not part of the solution sought, which we could compare to what we consider as noise. Indeed, as the number of external nodes, and thus the noise, increases, the solver is unable to discern what is part of the independent set from what is not. The most probable explanation is that as the noise in the data increases, the minimum energy gap decreases, leading the Hamiltonian to transition to more excited states, thus preventing the solver from identifying a solution to the problem.

This result introduces a new level of analysis to quantum computing. The power of these methods lies in the fact that, irrespective of the size or quality of the input, the solving time remains constant. However, we have demonstrated, at least for the maximum clique problem and its annexes, that this is not true for the quality of the solution, which is an equally important factor. Furthermore, we have demonstrated that the transition to excited states is not solely influenced by thermal or magnetic fluctuations that may affect the hardware. Rather, it is also influenced by the structure of the input itself, which may be noisy.

\subsection{Density}

The preceding experiment yielded a number of different tests, including the graph density test. As previously stated, the creation function enables the user to alter the density of the arcs, thus allowing the generation of a variety of graphs with differing densities.

A preliminary experiment was conducted by generating 36 graphs, each with a final dimension of 100 nodes, divided into three groups. Each group exhibited varying ratio level and was divided into four subgroups based on the density level. For the 40/60 and 50/50 ratio groups, there were no significant differences between the various density values. However, the 30/70 group demonstrated a decreasing trend in the average results as the density decreased, although ANOVA did not show any significant differences between subgroups.

Subsequently, another 150 graphs comprising 100 nodes each were generated, with a ratio of 0.42, divided into groups of 30 graphs based on density levels. Having extracted the solutions, a comparison was made between the various means of result, which did not show statistical trends, contradicting previous results. However, the analysis did reveal an intriguing phenomenon: for graphs with a density value below 15\%, the solver encounters significant difficulties in returning a solution, even a partial one. In fact, the number of null solutions for the various density levels is typically below 30\%, while at 15\% density it rises to 50\%, a value that is clearly unacceptable as it increases the risk of obtaining a solution equal to zero and consequently necessitates the repetition of the experiment.

This conclusion is also interesting, although less relevant than the ratio, as it allows a second value on which to make a preliminary analysis on the graphs. However, it is necessary to specify that the aforementioned conclusion, namely that it is desirable not to go below 15\% density, only covers graphs with a ratio of 0.42. It would be beneficial to generalise this theory by carrying out more in-depth experiments on other ratio classes.

\subsection{Connectivity Indices}

Another analysis was conducted using graph indices. The 30/70 ratio was employed to generate graphs with a solution and others with a null result. Several connectivity indices were calculated for these graphs, including mean and variance of vertex degrees, minimum and maximum eccentricity, centre and periphery size, centralisation, mean and variance of closeness, and mean and variance of betweenness. Finally, the differences between the indices of the graphs with a solution and those of the graphs without a solution were calculated.

The observed differences are relatively minor and do not justify further statistical investigation. It is important to note that the number of generated graphs is limited, which may not fully capture the underlying phenomenon. To address this, we conducted a second experiment, generating 100 graphs, with the aim of predicting the size of the result extracted by the quantum solver using classical learning models (SVM and KNN). This decision is based on the observation that learning models are occasionally capable of extrapolating patterns from data that appear to lack any discernible structure. The average accuracy of the two models is approximately 65\%. However, it should be noted that this result is likely influenced by a bias introduced by the asymmetry of the classes. This factor is due to the capability of the D-Wave solver. In fact, the majority of instances belong to the best result classes because in calculating the solution, the solver seeks the optimum. In order to balance the classes, it would be necessary to create specific instances, perhaps with a low ratio and/or density value. However, this would introduce an additional bias due to the difference between the graphs in the various classes. Indeed, the creation of specific instances would result in the model considering only the ratio and density variables as discriminants for the classes, thereby ignoring the index variables.

From this analysis, it can be preliminarily concluded that there is no apparent correlation between the indices describing a graph and the result extracted by the D-Wave solver. However, it was observed that training predictive models using connectivity indices is challenging, as it necessitates the generation of specific instances for training, which may introduce bias into the results. Nevertheless, further in-depth investigations could potentially identify subtle but significant differences for the solver.

\subsection{Number of Cliques}

Another factor that was subjected to testing was the influence of the number of cliques. A total of 33 graphs were generated, with a ratio of 0.42. These were divided into three categories: 11 graphs with one clique, 11 graphs with two cliques and 11 graphs with three cliques. It should be noted that the ratio value was calculated as follows:
\begin{equation}\label{eq:num_cri_for}
    R = \frac{C_m}{D_g - C_m}
\end{equation}
with $C_m$ maximum clique size and $D_g$ graph size. This is because we assume that the solver, by minimising the Hamiltonian, only searches for one clique and consequently that the other nodes act as noise, regardless of whether or not they belong to another clique.

Upon analysis of the solver results, an intriguing phenomenon was observed. The average of the solutions according to the number of cliques exhibited a relatively consistent trend across the three groups, whereas the sample variance exhibited a notable decline as the number of cliques increased. Between the variance of the results of the group with one clique and that with two, there was a 15\% decrease, while between two and three, there was a 50\% decrease. This result is noteworthy because it indicates that as the number of cliques increases, and thus density, the solver returns more homogeneous results.

To verify this hypothesis, we proceeded to calculate the variance values of the results of the density experiment for the 30/70 graphs, in which the lower limit of density was extrapolated to 0.15. Our findings indicate that as the density increases, the solver tends to return results with less variance, thereby producing solutions within a narrower range.

This conclusion is of interest with regard to the objective of obtaining more precise solutions. In the event that we have a loosely connected graph, it is possible to increase the number of arcs without increasing the value of the chromatic number in order to increase the probability of obtaining a result closer to the optimum, or, in general, with fewer outliers than the expected average. It is evident that the introduction of random arcs may also result in the modification of the maximum cliques. However, it is possible to conduct a series of experiments by increasing the density in different ways in order to either confirm or enhance the result already calculated on the original graph, provided that it does not exceed the value of the chromatic number.

In conclusion, the number of cliques is an impossible factor to calculate if one does not know the structure of the graph. However, it does influence the density value, which does not directly affect the quality of the solution. Instead, it affects the dispersion of the results with respect to an expected value.

\subsection{Size of the Graph}

As a final experiment, we sought to ascertain the solver's behaviour in approaching the maximum embeddable dimension.
To this end, we generated 24 graphs comprising one clique, six graphs with final size 121 and ratio 0.42, six graphs with size 143 and ratio 0.42, six graphs with maximum size (164) and ratio 0.42, and six graphs with maximum size and ratio 1.

The results indicated that the majority of the solutions were null, resulting from the solver's inability to extract solutions as the dimension increased, approaching the maximum embeddable dimension. The only noteworthy results were observed for graphs having dimension 164. For the ratio of 0.42, the solver extracts a clique of size 7, which represents approximately 14\% of the maximum solution. In contrast, for the ratio of 1, the solution size is 33, which equates to 40\% of the maximum clique. Consequently, the theory of solving improvement as the ratio of clique size to external node size increases is once again confirmed to be true.

In conclusion, the experiment demonstrated that in order to obtain reasonable solutions that are close to the optimum, it is optimal to query D-Wave with graphs of a maximum size of 100, although the number of maximum embeddable nodes are 164, otherwise there is a high risk of obtaining null or partial solutions that represent less than 40 per cent of the optimum solution.

\section{Conclusions and Directions for Future Work\label{sec:conclusions}}

In this paper, we analyses the solving ability of a quantum D-Wave solver for maximum cliques, an important NP-hard class problem. The objective is to understand how the graph structure can influence the solving capability of the quantum method.

Firstly, we have devised two algorithms. The first, for reducing the input size, decomposes the complementary graph in order to keep the maximum independent set unchanged, with a worst-case computation time of $O(n^5)$. The second algorithm generates benchmark graphs with controllable inputs in order to perform statistical analysis on the results.

In terms of the experiments conducted, we proceeded to test the significance of the ratio between clique size and nodes outside the clique for both the maximum clique problem and the minimum vertex coverage. We also tested the significance of various indices for describing a graph, including average degree, closeness, and so forth. Additionally, we tested the impact of density and the number of cliques on the solution. Finally, we tested the effect of an extreme input size (within the limits of embedding on the QPU) on the solution.

Subsequently, the majority of the results were subjected to statistical analysis, employing several techniques.
The findings yielded the following conclusions:
\begin{itemize}
    \item The ratio of clique size to the number of external nodes of 0.42 represents a threshold beyond which the solver's quality of solution begins to deteriorate.
    \item The density of the graphs does not appear to exert a significant influence on the size of the final solution, with the exception of graphs exhibiting a ratio of 0.42. In this instance, a density value below 15\% increases the probability of obtaining null solutions.
    \item Graph indices are not a reliable means of distinguishing different classes of results with the same ratio and/or size. Furthermore, training a predictive model with connectivity indices is challenging.
    \item The number of cliques is not a significant factor for the solver, as searching for a single clique considers all nodes not belonging to the solution, including those in other cliques, as external nodes.
    \item For the time being, the number of nodes that can be incorporated is limited to 164. As the quantum method approaches this limit, the probability of it returning null values increases.
\end{itemize}
In general, it can be concluded that there are various factors that influence the size of the solution returned by the quantum method and this analysis provides a valid starting point for sensible decomposition or restructuring of the problem, given the objective to be achieved.

Building on this, the work could be extended by a more detailed analysis of the importance of density and graph indices. In addition, it would be interesting to investigate whether a similar conclusion can be drawn for problems other than maximum cliques, and to determine what can be considered noise for other types of NP-hard problems. This would help to determine whether the theory of the solution-noise ratio can be generalised or if it should only apply to problems on graphs. Finally, it would be highly beneficial to continue this type of analysis in the future, when hardware may allow the incorporation of larger instances, in order to confirm or deny the conclusions here presented.

\printbibliography

\end{document}